# PICO- AND NANOSECOND LASER ABLATION OF MIXED TUNGSTEN / ALUMINIUM FILMS

M. Wisse,[a] L. Marot,[a] R. Steiner,[a] D. Mathys,[b] A. Stumpp,[c] M. Joanny,[d] J. M. Travère,[d] and E. Meyer,[a]

[a] Department of Physics, University of Basel, Klingelbergstrasse 82, CH-4056 Basel, Switzerland

[b] Centre of Microscopy, University of Basel, Klingelbergstrasse 50/70, CH-4056, Basel, Switzerland

[c] University of Applied Sciences Northwestern Switzerland (FHNW), Institute of Product and Production Engineering, Klosterzelgstrasse 2, CH-5210 Windisch, Switzerland

[d] CEA, IRFM, F-13108 Saint-Paul-lez-Durance, France

Keywords: ITER; First mirrors; Laser cleaning

Corresponding author for proofs and for sending page charge invoice to:

Dr. Laurent Marot, University of Basel, Department of Physics, Klingelbergstrasse 82, CH-4056 Basel, Switzerland

Tel.: +41 61 267 37 20

Fax.: +41 61 267 37 84

laurent.marot@unibas.ch

11 pages text, 8 figures, 1 table.

# PICO- AND NANOSECOND LASER ABLATION OF MIXED TUNGSTEN / ALUMINIUM FILMS

M. Wisse,[a] L. Marot,[a] R. Steiner,[a] D. Mathys,[b] A. Stumpp,[c] M. Joanny,[d] J. M. Travère, and E. Meyer,[a]

**Abstract**

In order to extend the investigation of laser-assisted cleaning of ITER-relevant first mirror materials to the picosecond regime, a commercial laser system delivering 10 picosecond pulses at 355 nm at a frequency of up to 1 MHz has been used to investigate the ablation of mixed aluminium (oxide) / tungsten (oxide) layers deposited on poly- and nanocrystalline molybdenum as well as nanocrystalline rhodium mirrors. Characterization before and after cleaning using scanning electron microscopy (SEM) and spectrophotometry shows heavy dust formation, resulting in a degradation of the reflectivity. Cleaning using a 5 nanosecond pulses at 350 and 532 nm, on the other hand, proved very promising. The structure of the film remnants suggests that in this case buckling was the underlying removal mechanism rather than ablation. Repeated coating and cleaning using nanosecond pulses is demonstrated.

Ⅰ. INTRODUCTION

In ITER, each optical diagnostic system will rely on a number of mirrors to relay light from the plasma through the neutron shielding surrounding the machine. The most critical mirror is the one nearest the plasma, the so-called first mirror, which is expected to have a change of optical performance either due to erosion or the deposition of particles eroded elsewhere.[1,2]

An outstanding issue is how to remove deposits in situ, as physical replacement of the mirrors is deemed undesirable on account of the downtime incurred. One cleaning technique that may be considered for in situ application is laser cleaning, provided that the laser not be sensitive to the magnetic field inside the tokamak, which would depend on the type of laser and its location. Our previous paper[3] contained a review of laser-assisted cleaning experiments on mirrors, carried out in the fusion community to date. Since then, a number of additional experiments have appeared in the literature. Our finding that the results obtained using visible radiation for cleaning may be improved by additional exposure to UV radiation has been confirmed in Ref. 4, where an excimer laser operating at 193 nm was employed successfully to further improve the optical properties of mirrors after initial cleaning using an Nd:YAG laser (532 nm). Heat transfer calculations suggest a ~ 5 nanosecond laser pulse would be optimal for 100 nm Be film.[5] Moreover, a laser system (Q-switched Nd:YAG laser with a fundamental wavelength of 1064 nm and 7 nanosecond pulses) has been used to remove carbon deposits from rhodium mirrors.[6] The laser damage threshold for 3000 laser pulses lies in the interval between 400-550 mJ/cm$^2$ as measured in Ref. 3, which was also measured in vacuum.

In Ref. 7, the multi-pulse laser induced damage threshold was investigated experimentally using a Nd:YAG laser with a pulse duration of ~ 5 nanosecond in air and vacuum, showing that the value in vacuum was twice that in the air, as previously reported

As the laser cleaning mechanisms were described exhaustively in Ref. 3, a brief summary is presented here. There are a number of processes that can play a role in the removal of material from a surface by pulsed laser radiation, depending on the type of material and the duration, wavelength and fluence of the laser pulse. The first step is the absorption of the laser radiation by electrons in the material. For pulses in the sub-picosecond range, the material removal occurs mainly after the pulse and or through the process of thermal evaporation. In the case of a 5 nanosecond pulse, thermal diffusion during the laser pulse becomes important. In this case, the free electrons have the time to thermally equilibrate with the ion lattice during the pulse, resulting in a much larger heat affected zone than in the case of ultrashort pulses, where the heat affected zone is confined to the optical skin depth. The ablation process is now thermal, whereby the ions escape the surface when their kinetic energy exceeds the binding energy.

In this regime, the ablation threshold fluence is known to scale with the square root of the pulse duration,[8,9] a relation which breaks down for pulses in the picosecond range. A decrease of the threshold fluence for pulses shorter than a picosecond was reported in Ref. 10. In our previous paper[3] we reported on laser cleaning experiments using a nanosecond laser system to remove various types of deposits and suggested the possible superiority of picosecond pulses. The present work describes experiments using a commercial picosecond laser micromachining system to ablate metallic layers containing a mixture of aluminium and tungsten from polycrystalline molybdenum and rhodium surfaces.[11] Aluminium serves as a substitute for beryllium in this case, which will be present in ITER but cannot be handled in most labs due to its toxicity.[12]

Ⅱ. SAMPLE PREPARATION

The samples consisted of a set of six 25 mm diameter round mirrors. Five of these were stainless steel ones, of which three had been coated with a nanocrystalline rhodium layer with a thickness between 2.3 and 4.5 microns, while the other two had been coated with a 4.5 micron nanocrystalline molybdenum layer using magnetron sputtering.[11] The sixth sample was made of polycrystalline molybdenum. The samples were coated by adding a mixture of aluminium and tungsten to an RF deuterium/argon plasma using magnetron sputtering until the reflectivity in the visible, as measured by an in situ reflectometry system,[13] had dropped by about 30 %. This kind of porous oxidized coating, [12] as expected to be in ITER will lead to a severe degradation of the reflectivity, a loss of 30% being considered a serious concern for the affected diagnostic. Measurements of the deposited film thickness under similar conditions provide an estimate of the order of 30 nm for the present films. The argon/deuterium ratio was about 1:9 by partial pressure, at a total pressure of 0.032 mbar. During deposition the samples were heated to 150 °C and biased to -200 V. For more information on the plasma deposition setup see Ref. 3 and 14. Although the composition of the films as measured by X-ray photoelectron spectroscopy (XPS) (Ref. 15), turned out to vary between pure aluminium / aluminium oxide and pure tungsten, the results proved very similar. The morphology of the films was investigated by SEM (Hitachi S-4800 field emission at 5 kV).

## Ⅲ. CLEANING EXPERIMENT

The experiments were carried out at the University of Applied Sciences and Arts Northwestern Switzerland using an industrial picosecond laser micromachining system built around the Duetto laser from Time Bandwidth Products, delivering 10 picosecond pulses at 355 nm at a frequency up to 1 MHz, with a Gaussian spot size of 30 microns at full width at half maximum (Fig. 1 a), b)). The samples were mounted in a vacuum chamber during the experiment (Fig. 1 c)), with the laser beam coming in from above. The chamber was evacuated using a conventional pumping system consisting of a primary pump and a turbo. Due to a faulty pressure gauge the pressure was not recorded. However, experience with the same system after these experiments provides a pressure range of $10^{-8}$-$10^{-7}$ mbar. An optical scanner was used to scan the laser beam across the sample surface, using a simple zigzag scan pattern whereby the beam was interrupted at the turning points to prevent burn-in. In order to identify a suitable pulse energy for the cleaning experiment, 4x4 mm squares on the first sample were exposed at subsequently lower energies, using a pulse repetition frequency of 1 MHz, until the damage incurred by the mirror surface seen to diminish by visual inspection. The starting energy was 0.71 J/pulse and the lowest energy was 0.03 J/pulse. A new area was used for each energy. The scanning speed was 1500 mm/s and the spatial pulse overlap was 97 %, both in the x- and y-direction, resulting in ~$10^3$ shots being fired at each location, which is similar to the experiments reported in Ref. 3. Damage to the substrate material (in the 4x4 mm$^2$ area) was seen to occur down to about 0.2 µJ per pulse. Having established a suitable parameter range using the first sample, a number of larger patches were cleaned on the remaining samples. Different patches were subjected to different numbers of exposures and pulse energies. Visual inspection suggested that the films had been removed from the surface at 0.07 µJ per pulse, lower than the damage threshold of the underlying substrate material.

Finally, two patches on the last sample were cleaned using nanosecond system at 532 and 350 nm,[3] in order to make a comparison with those areas exposed using the picosecond system. The exposures using the picosecond system were done with the samples lying horizontally, while the exposures using the nanosecond system were done with the samples mounted vertically.

IV. **RESULTS**

The results for all samples used in the experiment were very similar and to illustrate the results we will consider a polycrystalline molybdenum sample coated with aluminium and aluminium oxide, of which one half was cleaned with the picosecond and the other half with the nanosecond laser, see Fig. 2. Figure 3 shows SEM images corresponding to Fig. 2. The left two areas were cleaned using the picosecond laser. Area #2 was exposed to 0.13 µJ/pulse and area #1 to 0.07 µJ/pulse for $10^3$ pulses. Dust may be seen in particular in area #1. The two images on the right are of areas that were cleaned using the nanosecond laser, area #3 being exposed at 532 nm and #4 at 350 nm. Most of the film has disappeared from #4, though some patches are still left. The structure of the film remnant and the absence of small dust particles suggest that buckling was the removal mechanism. The SEM images in Fig. 4 correspond to location #5 in Fig. 2. They show an area adjacent to a cleaned area, before (a) and after (b) cleaning with the picosecond laser, showing dust has spread from the cleaned area during the cleaning process. X-ray energy dispersive spectrometry (EDS) measurements (not shown here) revealed that the dust is consists mainly of aluminium, as might be expected. XPS measurements also show that the molybdenum core level was detected on the two patches cleaned with the nanosecond system, but not on two patches cleaned with the picosecond system. Also, the dust was firmly attached to the surface, remaining in place while handling the samples after the experiment. The reflectivity of each of the four patches before and after cleaning is compared to the reflectivity before coating in Fig. 5. The procedure that was used for measuring the reflectivity is described in detail in Ref. 16. Both patches cleaned with the picosecond system show a decrease of the reflectivity, attributable to dust formation, while those cleaned with the nanosecond laser both show a near complete recovery. Finally, Table Ⅰ lists the atomic concentrations before and after cleaning as determined by XPS. The molybdenum core level was measured in the areas cleaned with the nanosecond system, showing the removal of the coating. Note that although no tungsten was measured by XPS on the surface after coating, it is found deeper inside the coating even if both magnetrons were used during the deposition. Another sample nicely illustrates the increasing oxidation of tungsten with an increasing number of exposures. Fig. 6 shows four XPS measurements of the W4f core level, taken after different numbers of exposures using the picosecond laser. Each spectrum has been normalized to one. It is quite likely that each laser pulse produced a short temperature rise and hence the oxidation process. The concentration of tungsten oxide relative to metallic tungsten increased with the number of exposures.

The results for all samples used in the experiment were very similar and may be summarized as follows:

- Severe dust formation is observed when using picosecond pulses, covering the cleaned as well as the surrounding areas.
- Treatment with the picosecond laser did not result in complete removal of the deposit.
- A reduction of the reflectivity was observed for all areas treated with the picosecond laser. Enhancement of the reflectivity was observed for the two areas cleaned with the nanosecond laser.
- Increasing oxidation of W with increasing energy / number of exposures to picosecond pulses.
- The film has all but disappeared from the areas treated with the nanosecond laser, without the formation of dust. The structure of the remnants suggest that buckling was the removal mechanism.

## Ⅴ. EXAMPLE OF REPEATEAD CLEANING

As a final example of nanosecond cleaning we shall briefly consider a polycrystalline molybdenum mirror that was coated and cleaned using successive exposures at 532 and 230 nm, a technique that was shown to enhance dust removal in Ref. 3. The pulse energy was 1 mJ/pulse for 1000 pulses. The speed depends on the spot size of the beam and the desired overlap between consecutive pulses, taking into account that the laser fires at 20 Hz. The spot size of the beam chosen was 300 µm, with an overlap of 250 µm, meaning that the sample travelled 50 µm every 50 ms, i.e. with 1 mm per second. The overlap exists for both the x- and y-direction, so that in this case the sample moved by 50 µm in the y-direction after each scan line in the x-direction. The dimension of the cleaned area was 17 × 17 mm, resulting in a total time of about 1.5 h per cleaning cycle. After the second cleaning the complete process was repeated to simulate cycles of coating and cleaning. The applied coating consisted of an Al/Al oxide/W mixture with an Al/W ratio of 9:1 and an estimated total thickness of the order of 30 nm. The sample is shown in Fig. 7. The specular and diffuse reflectivities are shown in Fig. 8, showing a substantial recovery of the reflectivity (a) and a negligible increase of the diffuse reflectivity (b).

## VI. CONCLUSIONS

A comparison has been made between pico- and nanosecond ablation of mixed aluminium/tungsten films. Severe dust formation and incomplete removal of the coatings are seen to prevent recovery of the reflectivity. It may be possible to remove the dust by irradiation with nanosecond pulses, though this remains to be investigated. However, nanosecond pulses proved to be efficient in removing the coating, be it with the sample in a vertical rather than a horizontal orientation, which may also affect the amount of accumulated dust. Nevertheless, the structure of the film remnants as shown in Ref. 3 suggests that buckling was the removal mechanism, rather than direct ablation as is the mechanism in the case of picosecond pulses. This may very well reduce the amount of dust formed. It would be useful to investigate the influence of the sample orientation, in particular for the picosecond regime. As it stands, however, the picosecond system does not offer an advantage over the nanosecond system, which has now demonstrated the removal of various types of coatings from various surfaces, as well as repeated removal and recovery of the reflectivity.

**ACKNOWLEDGMENTS**

The authors would like to thank the Swiss Federal Office of Energy and the Federal Office for Education and Science for their financial support. This work was also supported by the Swiss National Foundation (SNF) and the National Center of Competence in Research on Nanoscale Science (NCCR-Nano). Swissphotonics is gratefully acknowledged for its financial support.

TABLE Ⅰ: Atomic Concentrations on the Surface as Determined by XPS

| Location | Al (%) | W (%) | O (%) | C (%) | Mo (%) |
|---|---|---|---|---|---|
| Coating surface | 36 | 0 | 64 | 0 | 0 |
| #1 (ps) | 14 | 0 | 26 | 60 | 0 |
| #2 (ps) | 12 | 0 | 22 | 66 | 0 |
| #3 (ns) | 21 | 2 | 44 | 11 | 22 |
| #4 (ns) | 14 | 3 | 34 | 19 | 30 |

List of figures:

Fig. 1 (a, b) The setup, showing the vacuum chamber in position. Note that the exposures took place with the samples lying horizontally. (c) View of the samples inside the vacuum chamber through the quartz window, after the cleaning experiments. The cleaned patches are clearly visible on the mirrors.

Fig. 2. Polycrystalline molybdenum mirror, the two patches on the right were cleaned using the nanosecond system. The patches on the left, the bottom one of which is particularly difficult to make out, were cleaned using the picosecond system. For each patches $10^3$ pulses were used.

Fig. 3. Surface view SEM images, the numbers correspond to Fig. 2. Images were taken in the middle of each cleaned area.

Fig. 4. SEM images correspond to location #5 in Fig. 2. Area adjacent to a cleaned area, before (a) and after (b) cleaning with the picosecond laser, showing dust has spread from the cleaned area during the cleaning process.

Fig. 5. Specular reflectivity of each patch before and after coating, and after cleaning. The two patches cleaned with the picosecond system show a reflectivity lower than that of the coating due to the dust remaining on the surface. The two patches cleaned using the nanosecond system show an improved reflectivity (colour online).

Fig. 6. W4f core level XPS spectrum, measured on the coating and three cleaned areas with an increasing number of exposures with the picosecond system. The increase of the oxidized component relative to the metallic component is evident (colour online).

Fig. 7. (a) PcMo mirror that was coated and cleaned twice. The two SEM images are of the coating (b) and the cleaned area (c).

Fig. 8. Specular (a) and diffuse (b) reflectivity after two coating/cleaning cycles.

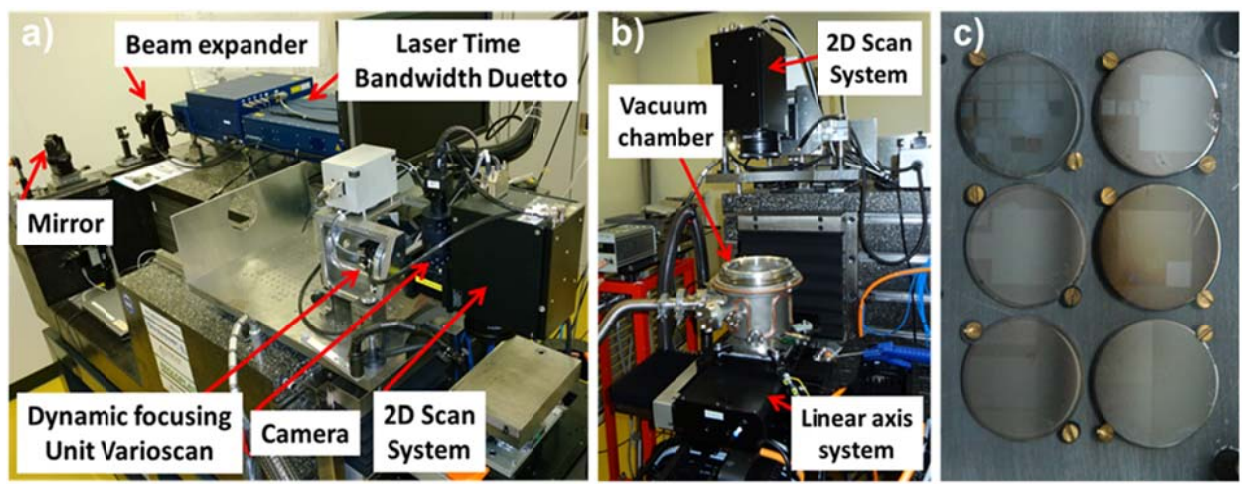

Fig. 1 (a, b) The setup, showing the vacuum chamber in position. Note that the exposures took place with the samples lying horizontally. (c) View of the samples inside the vacuum chamber through the quartz window, after the cleaning experiments. The cleaned patches are clearly visible on the mirrors.

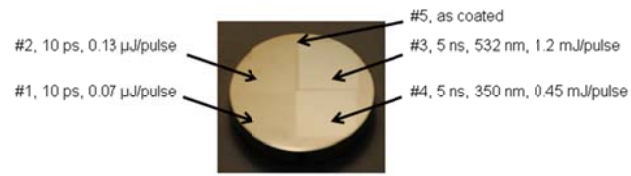

Fig. 2. Polycrystalline molybdenum mirror, the two patches on the right were cleaned using a nanosecond system. The patches on the left, the bottom one of which is particularly difficult to make out, were cleaned using the picosecond system. For each patches $10^3$ pulses were used.

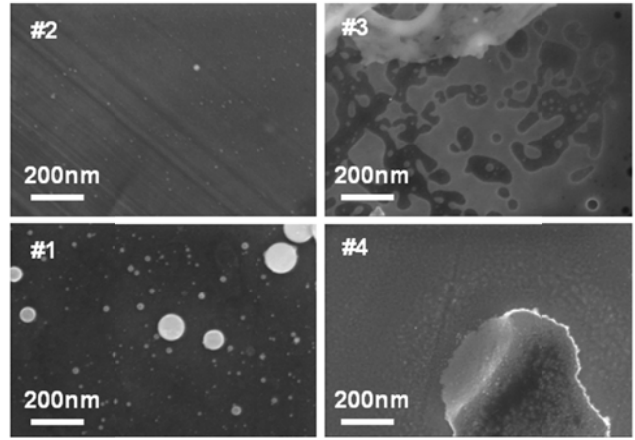

Fig. 3. Surface view SEM images, the numbers correspond to Fig. 2. Images were taken in the middle of each cleaned area.

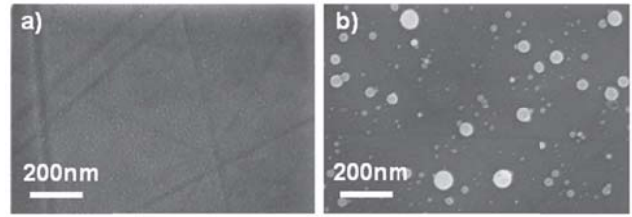

Fig. 4. SEM images correspond to location #5 in Fig. 2. Area adjacent to a cleaned area, before (a) and after (b) cleaning with the picosecond laser, showing dust has spread from the cleaned area during the cleaning process.

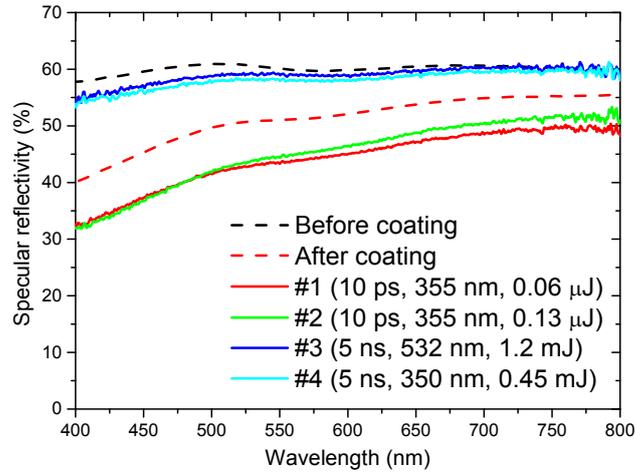

Fig. 5. Specular reflectivity of each patch before and after coating, and after cleaning. The two patches cleaned with the picosecond system show a reflectivity lower than that of the coating due to the dust remaining on the surface. The two patches cleaned using the nanosecond system show an improved reflectivity (colour online).

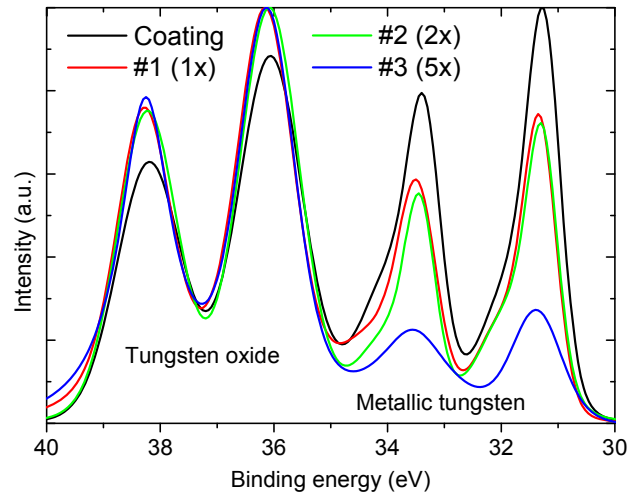

Fig. 6. W4f core level XPS spectrum, measured on the coating and three cleaned areas with an increasing number of exposures with the picosecond system. The increase of the oxidized component relative to the metallic component is evident (colour online).

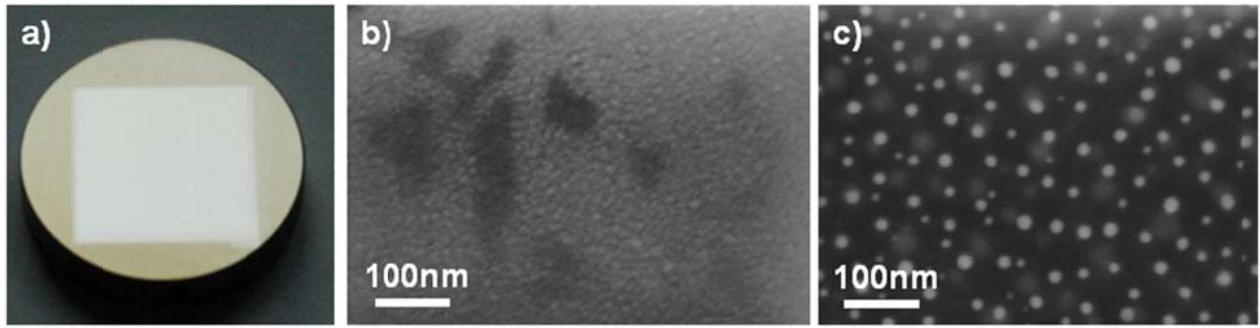

Fig. 7. a) PcMo mirror that was coated and cleaned twice. The two SEM images are of the coating (b) and the cleaned area (c).

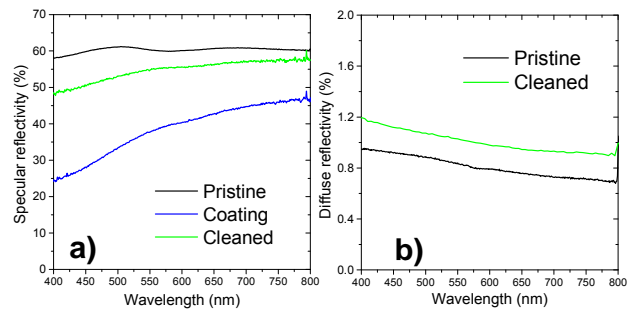

Fig. 8. Specular (a) and diffuse (b) reflectivity after two coating/cleaning cycles.

**REFERENCES**

1. A. LITNOVSKY et al., "Progress in research and development of mirrors for ITER diagnostics," *Nucl. Fusion*, **49**, 075014 (2009).

2. V. VOITSENYA et al., "Diagnostic first mirrors for burning plasma experiments," *Rev. Sci. Instrum.*, **72**, 475 (2001).

3. M. WISSE, L. MAROT, B. EREN, R. STEINER, and E. MEYER, "Laser damage thresholds of ITER mirror materials and first results on in situ laser cleaning of stainless steel mirrors," *Fusion Eng. Des.*, **88**, 388-399 (2013).

4. R. HAI et al., "Characterization and removal of co-deposition on the first mirror of HL-2A by excimer laser cleaning," *J. Nucl. Mater.,* **436**, 118-122 (2013).

5. C. H. SKINNER et al., "Laser cleaning of candidate diagnostic mirrors for ITER," *Fusion Science and Technology*, **64**, 1-7 (2013).

6. A. UCCELLO et al., "Laser cleaning of pulsed laser deposited rhodium films for fusion," *Fusion Eng. Des.*, **88**, 1347-1351 (2013).

7. S. KAJITA et al., "Enhancement of multi-pulse laser induced damage threshold on Cu mirror under vacuum condition," *Optics express*, **21**, 17275-17284 (2013).

8. A-C. TIEN et al., "Short-pulse laser damage in transparent materials as a function of pulse duration," *Physical Review Letters*, **82**, 3883-3886 (1999).

9. J. JIANG et al., "Femtosecond laser ablation: challenges and opportunities," Proceeding of NSF Workshop on Research Needs in Thermal, Aspects of Material Removal; Stillwater, Oklahoma, 163-177 (2003).

10. C. CURRAN et al., "Effect of wavelength and incident angle in the laser removal of particles from silicon wafers," 20th International Congress on Applications of Lasers and Electro-Optics, Jacksonville, (2001).

11. M. JOANNY, J. M. TRAVERE, S. SALASCA, L. MAROT, E. MEYER et al., "Achievements on Engineering and Manufacturing of ITER First-Mirror Mock-ups," *IEEE Trans. Plasma Sci.*, **40**, 692-696 (2012).

12. L. MAROT, C. LINSMEIER, B. EREN, L.MOSER, R. STEINER, and E. MEYER, "Can aluminium or magnesium be a surrogate for beryllium: A critical investigation of their chemistry," *Fusion Eng. Des.*, **88**, 1718-1721 (2013).

13. M. WISSE, B. EREN, L. MAROT, R. STEINER, and E. MEYER, "Spectroscopic reflectometry of mirror surfaces during plasma exposure," *Rev. Sci. Inst.*, **83**, 013509 (2012).

14. B. EREN, L. MAROT, M. WISSE, D. MATHYS, M. JOANNY, J.-M. TRAVERE et al., "In situ evaluation of the reflectivity of molybdenum and rhodium coatings in an ITER-like mixed environment," *J. Nucl. Mater.*, **438**, S852-S855 (2013).

15. B.EREN, L. MAROT, M. LANGER, R. STEINER, M. WISSE, D. MATHYS, and E. MEYER, "Progress in research and development of mirrors for ITER diagnostics," *Nucl. Fusion,* **51**, 103025 (2011).

16. M. WISSE, L. MAROT, A. WIDDOWSON, M. RUBEL, D. IVANOVA, R. P. DOERNER et al., "Laser cleaning of beryllium-containing first mirror samples from JET and PISCES-B," submitted to *Fusion Eng. Des.* (2013).